%% file: main.tex
\documentclass[a4paper,conference]{IEEEtran}

\usepackage{cite}
\usepackage{amsmath,amsfonts}
\usepackage{algorithmic}
\usepackage{algorithm}
\usepackage{array}
\usepackage[caption=false,font=normalsize,labelfont=sf,textfont=sf]{subfig}
\usepackage{textcomp}
\usepackage{stfloats}
\usepackage{url}
\usepackage{verbatim}
\usepackage{graphicx}
\usepackage{bbm}
\usepackage{dsfont}
\usepackage{xcolor}
\usepackage{csquotes}

\def\BibTeX{{\rm B\kern-.05em{\sc i\kern-.025em b}\kern-.08em
    T\kern-.1667em\lower.7ex\hbox{E}\kern-.125emX}}

\begin{document}

\title{Unsupervised Graph Neural Network Framework for Balanced Multipatterning in Advanced Electronic Design Automation Layouts}

\author{%
\IEEEauthorblockN{%
  Abdelrahman Helaly\IEEEauthorrefmark{1}\IEEEauthorrefmark{2},
  Nourhan Sakr\IEEEauthorrefmark{2},
  Kareem Madkour\IEEEauthorrefmark{1},
  Ilhami Torunoglu\IEEEauthorrefmark{1}
}
\IEEEauthorblockA{\IEEEauthorrefmark{1}Siemens EDA}
\IEEEauthorblockA{\IEEEauthorrefmark{2}The American University in Cairo, Cairo, Egypt}
}

\maketitle 

\begin{abstract}\input{Abstract}
\end{abstract}

\section{Introduction}
\input{Intro}

\section{Literature review}
\label{sec:lite}
\input{Lite}

\section{Problem Formulation}
\label{sec:prob}
\input{problem}

\section{Methodology}
\label{sec:method}
\input{method}

\section{Initial solution generation}
\label{sec:init}
\input{init}

\section{Refinement}
\label{sec:refine}
\input{refine}

\section{Experiments}
\label{sec:exp}
\input{exp}

\section{Future work}
\label{sec:fut}
\input{future}

\section{Conclusion}
\input{conc}

\bibliographystyle{ieeetr}
\bibliography{my_library}

\end{document}

%% file: Abstract.tex
Multipatterning is an essential decomposition strategy in electronic design automation (EDA) that overcomes lithographic limitations when printing dense circuit layouts. Although heuristic-based backtracking and SAT solvers can address these challenges, they often struggle to simultaneously handle both complex constraints and secondary objectives. In this study, we present a hybrid workflow that casts multipatterning as a variant of a constrained graph coloring problem with the primary objective of minimizing feature violations and a secondary objective of balancing the number of features on each mask. Our pipeline integrates two main components: (1) A GNN-based agent, trained in an unsupervised manner to generate initial color predictions, which are refined by (2) refinement strategies (a GNN-based heuristic and simulated annealing) that together enhance solution quality and balance. Experimental evaluation in both proprietary data sets and publicly available open source layouts demonstrate complete conflict-free decomposition and consistent color balancing.
The proposed framework provides a reproducible, data-efficient and deployable baseline for scalable layout decomposition in EDA workflows.

~\\
\begin{IEEEkeywords}
Electronic design automation; Graph neural networks; Multipatterning; Layout decomposition; Design for manufacturability; Machine learning; Graph Coloring.
\end{IEEEkeywords}

~\\
{\small \spaceskip=3sp Correspondence:  abdelrahman.helaly@siemens.com, n.sakr@columbia.edu}

%% file: Intro.tex
As semiconductor technology progresses, the demand for higher circuit densities continues to surpass the limits of conventional lithographic techniques. The ongoing reduction in feature size introduces increasingly complex manufacturing constraints, making it difficult to accurately print intricate patterns on a single mask without defects. To address these challenges, modern electronic design automation (EDA) tools and fabrication processes rely on \textit{multipatterning}, which is a layout decomposition technique that ensures manufacturability while preserving design integrity.
In modern integrated circuit (IC) design, \textit{Design Rule Checking} (DRC) is a critical step that ensures that the physical layout complies with a set of rules derived from the manufacturing constraints. These rules include the requirements on spacing, width, enclosure, and other geometric and connectivity constraints. A \textit{DRC violation} occurs when any layout feature fails to satisfy one or more of these design rules. An example is a spacing violation that occurs when metal lines or vias are too close to each other.
These manufacturing constraints are particularly critical when adjacent features are placed too close, resulting in unintended merging during the printing process. Such violations are typically detected through Design Rule Checking (DRC) decks. multipatterning mitigates these issues by decomposing the layout into multiple masks and systematically separating the features that would otherwise conflict. Each mask is printed in a distinct lithographic step, and the final pattern emerges from the combination of these layers.

\subsection{Constrained Graph Coloring in EDA}
Depending on the manufacturing constraints and feature density, multipatterning techniques are commonly categorized into double patterning (DP), triple patterning (TP), and quadruple patterning (QP), which correspond to the number of masks used. Assigning features to masks presents a computationally difficult problem that maps directly to a constrained version of graph coloring, which is NP-hard for \( k \geq 3 \), where \( k \) represents the number of colors (masks).

In practice, multipatterning presents challenges that extend beyond the classical graph coloring. Manufacturing rules often impose layout-specific constraints. For example, certain anchor features may be required to share the same mask, and mask assignments must maintain a balanced distribution of features to ensure uniform processing.

To address these domain-specific constraints, specialized algorithms  have been developed that build upon and extend conventional graph-coloring techniques. Although the underlying problem remains computationally difficult, layout simplification methods, such as feature clustering or conflict graph reduction can help reduce the problem size and improve runtime efficiency.

\subsection{Multipatterning Algorithms in Industry and Their Limitations}
Traditional approaches to constrained graph coloring in EDA predominantly depend on heuristic-based methods such as greedy algorithms and constraint solvers (SATs). Although these methods can perform well for small or moderately constrained scenarios, they often require extensive manual tuning and are difficult to adjust as constraints grow more complex or evolve over time. Moreover, SAT solvers are not well-suited for handling multi-objective functions, particularly when the solution surface is nonconvex, such as in cases that require satisfying all constraints and balancing nodes between masks.

Alternatively, learning-based approaches have gained traction recently. Techniques such as graph neural networks (GNNs), reinforcement learning (RL), and differentiable optimization offer the ability to capture intricate layout patterns and dynamically adapt to varying constraints. These models show promise in automating mask assignment decisions, reducing manual effort, and improving the solution quality under real-world manufacturing conditions. However, the application of these models introduces new challenges in stability, scalability, and training complexity, which necessitates careful management.

\subsection{Our Contribution}

To address the limitations of recent learning-based approaches, we propose a robust AI workflow that enables both accurate and balanced mask assignment for complex layouts. To the best of our knowledge, this work is the first to apply unsupervised Graph Neural Networks (GNNs) to this specific Electronic Design Automation (EDA) problem. Our approach reduces mask conflicts and minimizes the need for manual tuning, thereby contributing to more efficient and reliable manufacturing processes.

Our key contributions are as follows:
\begin{enumerate}
    \item We develop a novel GNN-based coloring method that operates in an unsupervised manner (without relying on labeled data) and is capable of handling complex, domain-specific constraints commonly encountered in EDA multipatterning problems.
    \item We propose a refinement strategy that integrates GNN-based heuristics and simulated annealing to correct violations and achieve balanced, conflict-free solutions across diverse layout scenarios.
\end{enumerate}

The rest of this paper is organized as follows.  In Section \ref{sec:lite} we survey existing graph-coloring approaches and group them according to their underlying learning paradigms.  In Section \ref{sec:prob}, we then formally define the graph-coloring problem and introduce the notation used throughout the paper.  Section \ref{sec:method} gives an overview of our proposed framework, after which Section \ref{sec:init} dives into the specifics of our initial-solution generation strategy.  Building on that, Section \ref{sec:refine} describes how we iteratively refine these solutions to improve the color quality.  In Section \ref{sec:exp}, we present a comprehensive set of experiments that demonstrate the effectiveness of our approach on a variety of benchmarks.  Finally, Section \ref{sec:fut} concludes with a summary of our contributions and outlines promising directions for future research.

%% file: Lite.tex
The multipatterning problem in EDA can be viewed as a constrained variant of the classical graph coloring problem. Unlike traditional graph coloring, where the objective is to minimize the number of colors required, multipatterning focuses on assigning colors in a way that eliminates design rule violations while also meeting additional requirements such as balancing across color groups.
The literature is rich with works that leverage various machine learning techniques for graph coloring for instance

H. Lemos et al. \cite{gcmeetdp} introduce GNN-GCP, a supervised graph neural network tailored for graph coloring. The authors employ a constraint satisfaction problem solver to generate the training data and subsequently employ GNNs, RNNs, and MLPs to determine the $k$-colorablity of a given graph. However, effectiveness and efficiency of supervised approaches heavily hinge on the availability of large, labeled training datasets comprising previously optimized instances of intricate problems \cite{erdos,potts}. Notably, the graph coloring problem is inherently defined by its constraints and does not require a pre-labeled dataset to begin with. Given the rapidly evolving nature of the EDA domain, such datasets may quickly become obsolete, making supervised methods less reliable for long-term applications. 

Schuetz et al. \cite{potts} propose an unsupervised GNN-based approach for graph coloring, drawing inspiration from the statistical physics Potts model. Their loss function penalizes edges between nodes sharing high-probability assignments to the same color, effectively promoting distinct color assignments. Building upon this GNN foundation, the approach naturally extends to weighted graph coloring through edge weight integration in the loss function or additional loss terms for specific industry needs.

Li et al. \cite{HK_Group} propose a margin‐based loss function that enforces a predetermined Euclidean distance between the predicted color embeddings of adjacent nodes. By removing the need for labeled training data, this approach and earlier ones offer major practical benefits. However, one-shot solution generation, where the model produces a complete coloring in a single inference, remains inherently unreliable. As the number of interdependent constraints increases, the stochastic characteristics of machine learning models exacerbate the accumulation of errors, leading to degraded performance in complex graph-coloring tasks.

Alternatively, Reinforcement learning (RL) has also been applied to the graph coloring problem, with a focus on sequential decision-making strategies. Watkins. et al. \cite{RL1} propose a Deep Q-learning network that sequentially selects the next node to be colored before applying an optimal rule-based coloring algorithm. Their primary objective is to minimize the number of colors used. However, in the EDA industry, the maximum number of colors is typically predefined. Consequently, the focus shifts from minimizing the number of colors to assigning them in a way that reduces design violations. Applying this approach in EDA would require developing additional heuristics within the rule-based component, which may not provide significant improvements over existing methodologies. More recently, Cummins et al. \cite{RL2} experiment with both Proximal Policy Optimization (PPO) and Deep Q-Network (DQN) to simultaneously select the next node and determine its color assignment. Their results indicate that PPO significantly outperforms DQN in this task. Nevertheless, training such RL agents remains challenging due to slow convergence and sample inefficiency,
 
As a takeaway, Supervised learning offers optimal training stability and safety but mandates the availability of datasets containing optimal constraint satisfaction solutions. Although unsupervised learning and knowledge distillation frameworks can alleviate this requirement, existing literature overlooks critical EDA domain constraints, particularly color balancing considerations that could be integrated into graph coloring optimization objectives.

%% file: problem.tex
The multipatterning problem in EDA can be viewed as a variant of the \textit{classical graph coloring problem} with additional constraints. While traditional graph coloring focuses on determining the minimum number of colors required to color a graph (the chromatic number), the multipatterning problem shifts the objective. Here, the goal is not minimizing the number of colors, but rather assigning them in a way that reduces design violations. In particular, the coloring must satisfy all constraints derived from DRC, while also addressing additional requirements such as balancing the number of colors used across the layout.

To address this, we use the Calibre tool from Siemens EDA. This tool decomposes a full-circuit layout into a collection of constrained graphs. Each graph is then subjected to layout and graph reduction techniques, producing a set of minimized graphs whose union of constraints is equivalent to the design rules of the original layout. From this point onward, we take these final reduced graphs $\{G_i\}$ as our input.

\subsection{Graph Definition}
For a given reduced graph $G_i = (V_i, E_i)$, each node $v\in V_i$ corresponds to a layout feature, and each edge $(u,v)\in E_i$ encodes a design‐rule constraint (DRC) between these features. The full constraint set $\mathcal{C}$ is derived from the DRC deck and includes {different‐mask constraints} on edges, which forbid assigning the same mask to adjacent features, and {graph‐level constraints}, such as balancing the number of features per mask.
Other constraint types lie outside our scope and are masked from this formulation.

For each subgraph $G_i$, we formalize the multipatterning problem by defining a coloring function
\begin{equation}
\mathrm{color}: V_i \longrightarrow \{1,2,\dots,K\},
\end{equation}
where each vertex in $V_i$ is mapped to one of $K$ distinct pattern types.

where $K$ is the maximum number of available masks. In practice, $K$ maybe lower than the chromatic number\footnote{The chromatic number is the minimum number of colors required so that no two adjacent nodes share the same color.}, so some DRC violations are inevitable.

\subsection{Objective Functions}
We define the mathematical foundation of our approach through a multi-objective optimization framework. Our formulation integrates two primary minimization objectives that must be addressed simultaneously to achieve effective mask partitioning. The first objective focuses on minimizing the number of edge conflicts shown in equation \ref{eq:coloring}. The second objective function targets minimizing the deviation from the ideal per-mask node distribution shown in equations \ref{eq:balance} and \ref{eq:balance_harsh}

\paragraph{1. Minimize edge‐conflict violations}
\begin{equation}
\label{eq:coloring}
\min \;\Bigl|\{(u,v)\in E_i \;\mid\; \mathrm{color}(u)=\mathrm{color}(v)\}\Bigr|.
\end{equation}
where $E_i$ represents the edge set of graph $G_i = (V_i, E_i)$

\paragraph{2. Balance node counts across masks}  
Let 
\[
n_c = \bigl|\{v\in V_i \mid \mathrm{color}(v)=c\}\bigr|
\quad\text{and}\quad
N = |V_i|.
\]
We penalize the deviations from the ideal per‐mask load $N/K$ by minimizing, for example,
\begin{equation}
\label{eq:balance}
\min\;\sum_{c=1}^K 
\Bigl(n_c - \tfrac{N}{K}\Bigr)^{2}.
\end{equation}
Alternatively, one can impose hard bounds

\begin{equation}
\label{eq:balance_harsh}
\bigl|n_c - \tfrac{N}{K}\bigr|\le\delta
\quad\forall\,c,
\end{equation}

for some tolerance $\delta$.

\paragraph{3. Combined formulation}  
In its most general form, we seek to minimize the total number of violated constraints over all $c_i\in\mathbf{C}$, where
\[
\mathds{1}\bigl(\mathrm{violation}(c_i)\bigr)
=
\begin{cases}
1, & \text{if constraint }c_i\text{ is violated},\\
0, & \text{otherwise},
\end{cases}
\]
so that
\begin{equation}
\min\;
\sum_{c_i\in\mathbf{C}}
\mathds{1}\bigl(\mathrm{violation}(c_i)\bigr).
\end{equation}


\begin{figure}
    \centering
    \includegraphics[width=\linewidth]{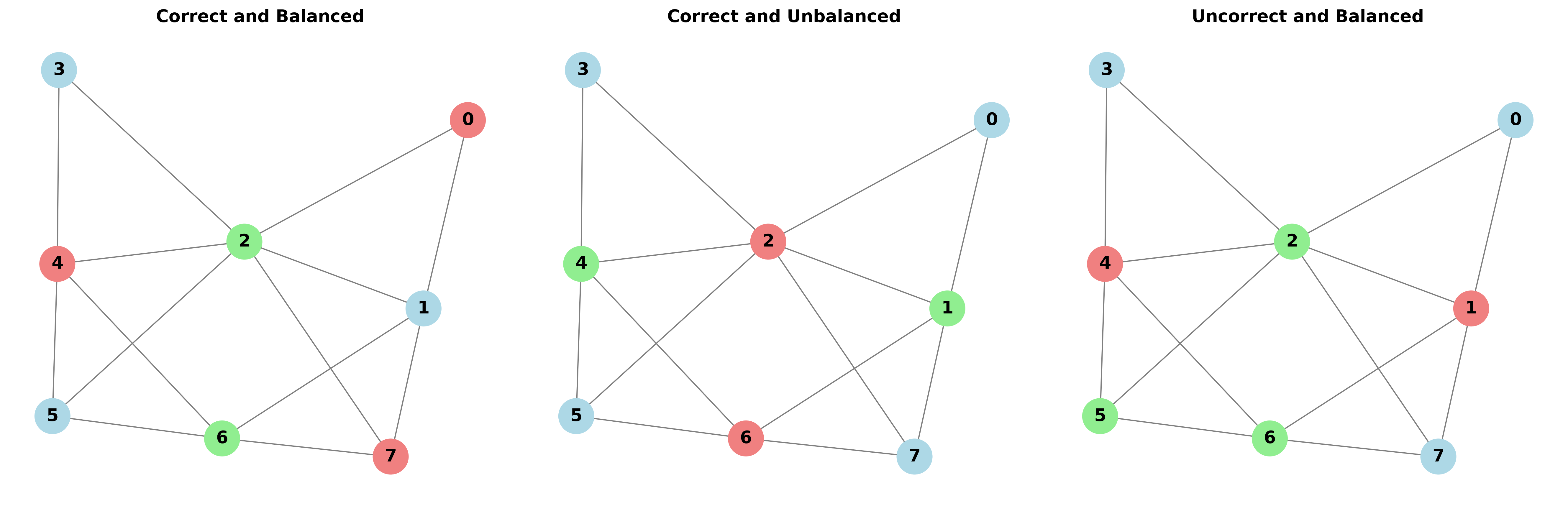}
    \caption{Comparison of valid and invalid graph colorings with varying balance. (Left:) A valid and balanced solution with partition sizes (3, 3, 2), where no adjacent nodes share the same color and the maximum partition size difference is 1. (Middle:) A valid but unbalanced solution with partition sizes (4, 2, 2), where the coloring constraint is satisfied but the maximum difference between partitions is 2. (Right:) An invalid but balanced solution with partition sizes (3, 2, 3) that achieves good balance but violates the graph coloring constraint. This illustrates the trade-off between satisfying hard constraints (valid coloring) and optimizing for balance.}
    \label{fig:exp_balance}
\end{figure}

%% file: method.tex
Multipatterning is a fundamental problem in layout verification, where circuit features must be assigned to different mask layers such that features on the same layer maintain sufficient spacing to ensure manufacturability. We utilize GNNs as a learning-based approach to solve multipatterning due to their ability to learn and exploit structural relationships within graphs, making them naturally suited for problems where local neighborhood information and global graph structure both influence optimal solutions.

\begin{figure}[t!]
  \centering
  \includegraphics[width=\linewidth]{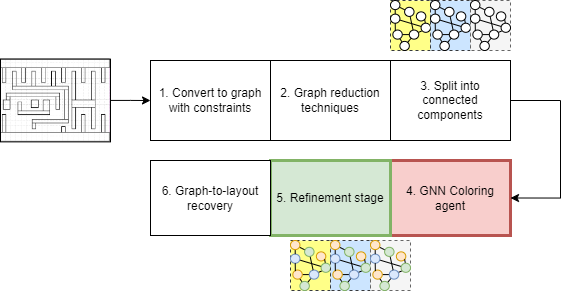}
  \caption{Methodology overview: for graph G, the coloring generation is divided into an initial solution (Step 4) and a refinement phase (Step 5).}
  \label{fig:pipe}
\end{figure}

We propose a two-stage design where (1) a learning-based solution is initially generated, before it moves to (2) an algorithmic-based refinement stage. Figure \ref{fig:pipe} illustrates our overall layout verification workflow and in this section, we break down the approach to each stage in full detail.

\subsection{Initial solution generation}
Our approach utilizes a layout-aware GNN agent that autonomously handles the initial coloring step. The agent primarily focuses on minimizing the number of violations, enabling effective graph coloring solutions even in cases where the problem is inherently infeasible \footnote{when the graph's chromatic number exceeds the number of colors available for the model to use}. Minimizing conflicts facilitates easier localization and resolution of design issues, even if later adjustments, such as increasing the number of masks (K) or modifying parts of the design, would arise.
Our GNN architecture specifically employs (1) {\it autonomous attention mechanisms} that dynamically identify and prioritize the most critical node relationships during the coloring process (2) {\it unsupervised learning} to dynamically learn the local and global graph structure without the need for labeled training data, enabling the model to adapt to diverse layout topologies; (3) {\it variable input adaptability} to support graphs of arbitrary sizes and structures without requiring fixed-size representations; and (4) {\it iterative message passing} that allows the agent to progressively refine its coloring solution through multiple rounds of constraint propagation and conflict resolution\footnote{These architectural principles are analogous to those employed in Large Language Models, where attention mechanisms, unsupervised learning, token-based adaptability, and iterative refinement enable effective sequence processing and generation.}. This iterative refinement process ensures that each coloring decision considers both immediate local constraints and broader structural dependencies across the entire graph layout.

\subsection{Refinement Stage}
Recognizing the probabilistic nature of machine learning outputs and their potential for error, we introduce a secondary refinement stage as a ``safety net'' for failed instances. This stage improves the coloring solution by integrating:

\begin{itemize}

     \item \textbf{GNN-Heuristic:} We loop through all nodes in an order determined by which nodes are most constrained (least number of correct colors left to color with) . For each node, the GNN provides its top \(N\) color predictions ranked by confidence. We then assign the highest-ranked color that satisfies both the local adjacency and global balance constraints before moving on to the next node.
    
    \item \textbf{Constraint Programming Solver (CSP) :} If violations remain, a CSP check ensures that the issue is due to uncolorability, not model limitations.
    
\end{itemize}  

The refined solution then undergoes a final balancing stage using simulated annealing, allowing color adjustments that enhance balance without introducing violations.

\subsection{Datasets}
We utilize two types of datasets to evaluate the proposed workflow

\subsubsection{Synthetic Data}
Using  Siemens's EDA toolkit, we generate synthetic graphs with varying node counts and a known chromatic number. These generated graphs are not random but are created with specific heuristics in mind to thoroughly test the in-house coloring engine.

\subsubsection{Real Data.}
\paragraph{Company data} We utilize an in-house training dataset of 2132 graphs and a test set of 608 graphs. All graphs were reduced from real representative layouts at Siemens. All graphs are unlabeled, which comes from the regression suite of the current rule-based tools.

\subsubsection{Open-source Datasets}

We utilize layout benchmarks provided by the OpenMPL framework~\cite{li2019openmplopensourcelayout}, focusing on representative designs derived from a scaled-down and modified version of the ISCAS benchmark suite. We use two sample layouts sim\_c1.gds and sim\_c2.gds.

%% file: init.tex
The proposed pipeline begins with the graph and feature specification, in which the structural properties of the input graph and the corresponding node and edge features are formally defined. This stage is followed by the development of a tailored GNN model, training protocol, inference procedure, and post-processing refinements.
This section introduces the learning component of the pipeline.

\subsection{Implementation Details}
To integrate an input graph into our framework, we define a structured initialization process that assigns feature representations to nodes and encodes constraints on the edges and at the global graph level.

Each input graph \( G = (V, E) \) is processed by associating every node \( v \in V \) with two complementary feature vectors: the Coloring Estimate Vector (\( F1 \)), which encodes the current coloring hypothesis, and the Node Embedding Vector (\( F2 \)), which captures structural and contextual information. During each propagation step, messages are exchanged between adjacent nodes based on their \( F1 \) representations. These messages are aggregated and used to update \( F2 \), which in turn informs the refinement of \( F1 \). This dual-update mechanism allows the network to disentangle structural context from solution-space exploration, while jointly optimizing both representations through learned transformations. These vectors collectively serve as the input representations for the GNN.

Each edge \( e = (u, v) \in E \) represents a hard constraint enforcing that the adjacent nodes \( u \) and \( v \) must be assigned different colors. These constraints are embedded into the message-passing mechanism of the GNN, guiding the exchange of information between connected nodes. The resulting coloring violations are penalized via a loss term denoted as \( L_1 \).

In addition to the node- and edge-level information, we incorporate global, graph-level constraints that influence the overall solution space. These constraints capture high-level properties of the graph-coloring problem such as balancing the number of nodes in each color, anchoring certain nodes to predefined colors, symmetry-breaking conditions, etc and are made accessible to the model as loss functions titled \(L_2, L_3, \dots, L_K\), each corresponding to a different global requirement.

This initialization stage ensures that both local (node and edge) and global structural information are made available to the model from the outset, thereby enabling context-aware learning throughout the training and inference process.

\subsection{Tailored GNN Layer}

The GNN layer in our model is specifically tailored to exploit the structural and constraint-based properties of the graph coloring problem. At its core, the layer performs iterative message passing to propagate information across the graph, enabling each node to refine its representation based on its local neighborhood and the coloring dynamics.

The proposed GNN model contains five custom message passing layers, designed to effectively process the graph and update the node coloring iteratively.

\begin{enumerate}
    \item \textbf{Message Passing:}  
    The graph is treated as an undirected graph, meaning a single edge allows messages to be sent in both directions. Various formulations of the message content were tested, and it was found that \( F1_{\text{src}} - F1_{\text{dst}} \) provides the most informative representation.  \footnote{Alternative message passing formulations were evaluated, including concatenation followed by multilayer perceptron processing, element-wise addition, and one-dimensional convolution.}
    Each message is scaled by an attention score computed as:  
    \begin{equation}
    \text{Attention Score} = \text{MLP}(\text{concat}(F1_{\text{src}}, F1_{\text{dst}}))
    \end{equation}
    This process implements a \textit{neighborhood attention} mechanism, where attention scores determine how much each neighbor's message contributes to the node's updated representation.
    \item \textbf{Message Aggregation:}  
    Messages from neighboring nodes are aggregated using different techniques, including maximum, minimum, mean and standard deviation.
    This aggregation captures the newly gathered neighborhood information, summarizing the local graph structure.

    \item \textbf{Node Embedding Update:}  
    The \( F2 \) vector, representing the node's internal embedding, is updated through an MLP. This MLP takes the previous \( F2 \) and the newly aggregated neighborhood information as input, enabling the model to incorporate both new knowledge and previously learned features. This effectively implements a form of \textit{self-attention} for controlling the influence of prior knowledge.

    \item \textbf{Coloring Estimate Update:}  
    The \( F1 \) vector, representing the current coloring estimate, is updated using another MLP. This MLP takes the previous \( F1 \) and the updated \( F2 \) as input, generating the next estimate for the node's coloring after normalization.
    
\end{enumerate}

\subsection{Loss Functions}
\label{sec:losses}

To guide the learning process, we design a suite of loss functions that address three core objectives: ensuring valid colorability, achieving color balance across nodes, and satisfying other global constraints. These losses are categorized into three groups: coloring losses (\( L_1 \)), balancing losses (\( L_2 \)), and other global constraint losses (\( L_3 \)).

\paragraph*{\textbf{Coloring Losses (\( L_1 \))}}

The primary objective of the model is to produce a valid coloring that satisfies the graph coloring constraints—namely, that adjacent nodes are assigned different colors. Several formulations of \( L_1 \) are used to penalize color conflicts and promote feasible assignments:

\begin{itemize}
    \item \textbf{Pairwise Loss}: Inspired by \cite{potts}, this loss penalizes adjacent nodes that share similar coloring probabilities. It is computed as the element-wise multiplication of the adjacency matrix and the dot product of node color probability vectors.
    \begin{equation}
        L_{\mathrm{pair}} = \sum_{(u,v)\in E} \bigl(p_u^\top p_v\bigr)
    \end{equation}

    \item \textbf{Clique Joint Probability Loss}: This term reduces the joint probability that all nodes in a clique (e.g., triangles for \( k=3 \)) are assigned the same color, thereby discouraging invalid clique colorings.
    \begin{equation}
    L_{\mathrm{clique}} \;=\; \sum_{C\in \mathcal{C}}
        \sum_{k=1}^K
    \prod_{v\in C} p_{v,k}
    \end{equation}

    \item \textbf{Unique Color Penalty}: 
    Given a set cliques \(\mathcal{G}=\{G_1,\dots,G_M\}\) 

    \begin{equation}
        L_{unique} \;=\;\sum_{G\in \mathcal{G}}
       \sum_{k=1}^K
         \Bigl\lvert\,\sum_{v\in G}p_{v,k} \;-\; 1\Bigr\rvert
    \end{equation}

\end{itemize}

\paragraph*{\textbf{Balancing Loss (\( L_2 \))}}

To promote equitable use of available colors across the graph, we define a balancing loss based on the discrepancy between the predicted color distribution and the uniform distribution. Let $p_{v,c}$ denote the soft assignment probability that node $v$ is assigned to color $c$, collected into the vector $p_v$
\[
\mathbf{p}_v = [p_{v,1}, \ldots, p_{v,K}]^\top, 
\quad \text{where } \sum_{c=1}^K p_{v,c} = 1.
\]
The average color usage distribution is
\[
\mathbf{u} = \frac{1}{|V|}\sum_{v \in V} \mathbf{p}_v,
\quad \mathbf{u} \in \Delta^{K-1}.
\]

We encourage $\mathbf{u}$ to align with the uniform distribution 
\[
\mathbf{u}^* = \Bigl[\tfrac{1}{K}, \ldots, \tfrac{1}{K}\Bigr]^\top
\]
by minimizing the Jensen–Shannon divergence:
\begin{equation}
    L_2 \;=\; 
    \text{JS}(\mathbf{u}\,\|\,\mathbf{u}^*) 
    \;=\; \tfrac{1}{2}\,\text{KL}(\mathbf{u}\,\|\,\mathbf{m})
        + \tfrac{1}{2}\,\text{KL}(\mathbf{u}^*\,\|\,\mathbf{m}),
\end{equation}
where $\mathbf{m} = \tfrac{1}{2}(\mathbf{u} + \mathbf{u}^*)$ is the mixture distribution. 
Minimizing $L_2$ encourages balanced utilization of the $K$ colors.

\paragraph*{\textbf{Anchoring Losses (\( L_3 \))}}

In cases where anchor nodes are known a priori—i.e., nodes with pre-assigned or class-specific colors—a cross-entropy loss is applied to encourage the model to preserve their labels. This supervision serves as an inductive bias, particularly useful in early training stages or in our two-stage learning procedure discussed below.

All losses are computed over soft probability distributions associated with coloring estimate vector (\( F1 \)), which are progressively refined throughout the training which will be described next.

\subsection{Training Procedure}
\label{sec:gnn_layer}

Training proceeds in two stages to balance feasibility and optimality across the defined objectives. While our model aims to satisfy both colorability and balance, we treat both constraints with a prioritized optimization strategy: colorability (\( L_1 \)) is the primary goal, and balancing (\( L_2 \)) is treated as a secondary objective that is optimized only when it does not significantly hinder feasibility.

\begin{enumerate}
    \item \textbf{Joint Initialization Phase:} In the initial phase, we optimize a combined objective \( L_1 + L_2 \), allowing the model to learn feasible and roughly balanced solutions. This provides a well-informed starting point for the subsequent focused training.

    \item \textbf{Priority-Guided Fine-Tuning:} In the second phase, we transition to a dynamically weighted objective \( L_1 + \beta L_2 \), where the coefficient \( \beta \) is adaptively adjusted. If improvements in balancing loss (\( L_2 \)) lead to deterioration in colorability (\( L_1 \)), we reduce \( \beta \) to refocus the model on feasibility. This adaptive mechanism ensures that the model prioritizes valid colorings while still promoting balance when possible.
\end{enumerate}

\paragraph*{\textbf{Joint training techniques}}

we experiment with several strategies for incorporating the balancing loss into the optimization process. The goal was to explore the trade-offs between feasibility and balance under different training dynamics. The strategies evaluated include:

\begin{itemize}
    \item \textbf{Dynamic Weighting:} The total loss is defined as 
    \begin{equation}
         L = L_1 + \beta L_2 
    \end{equation}
    , where \( \beta \) is adaptively adjusted based on the behavior of \( L_1 \), as described previously. This approach prioritizes feasibility while allowing controlled influence from the balancing term.

    \item \textbf{Single Optimizer with Gradient Reweighting:} Gradients are computed separately for \( L_1 \) and \( L_2 \), and then combined prior to the optimization step using a weighted sum:  
    \begin{equation}
        \nabla_\theta L = \nabla_\theta L_1 + \beta \nabla_\theta L_2,
    \end{equation}

    where \( \theta \) denotes model parameters. This technique allows fine-grained control over the contribution of each loss at the gradient level.

    \item \textbf{Dual Optimizer Scheme:} Each training instance is processed twice per step: once with respect to \( L_1 \) and once with respect to \( L_2 \). The model is updated using two separate optimizers, each with potentially distinct learning rates. This approach decouples the optimization dynamics of the two objectives and allows more flexible control of their learning trajectories.
\end{itemize}

\subsection{Inference Procedure}
\label{sec:itr_ref}
We guide the GNN convergence by employing an iterative refinement mechanism during inference to progressively enhance the solution quality. Our approach is conceptually analogous to iterative prompting in LLMs, where successive refinements of context progressively reduce errors or inconsistencies. Similarly, we start with a random color assignment to all nodes, followed by a forward pass through the graph. The resulting predictions update the initial \( F1 \) vectors, which become the input of the subsequent forward pass. Iterative updates to \( F1 \)  gradually improve the solution, converging towards a more accurate and feasible solution with each pass.

We experimented with various strategies for stopping criteria. Ultimately, employing a fixed number of iterations (\( N \)) proved to be simplest and most effective, delivering comparable results to more sophisticated termination methods.

%% file: refine.tex
\subsection{Correction Stage with GNN Heuristic}
\begin{algorithm}[!t]
  \caption{GNN Heuristic Refinement Block}
  \label{alg:passN_refinement}
  \begin{algorithmic}[1]
    \STATE {\bfseries Input:} 
      Predicted probabilities $\hat{P} \in \mathbb{R}^{N \times k}$, graph $G=(V,E)$
    \STATE {\bfseries Output:} 
      Final coloring $C \in \{1, \dots, k\}^N$
    \STATE Sort all nodes $v \in V$ by most constrained (least number of correct colors to color with), tie‐breaking by highest probability uncertainty
    \FOR{$i = 1$ \TO $N$}
      \STATE Let $v = \text{Node}[i]$
      \STATE Let $c^{*} = \arg\max_{c} \hat{P}_{v,c}$
      \IF{assigning color $c^{*}$ to $v$ causes no conflict in $G$}
        \STATE Assign $C[v] \leftarrow c^{*}$
      \ELSE
        \STATE Let $\mathcal{C}_{\text{safe}} = \{\,c : (v,c)\text{ causes no conflict}\,\}$
        \STATE Choose $c' = \arg\max_{c \in \mathcal{C}_{\text{safe}}} \hat{P}_{v,c}$
        \STATE Assign $C[v] \leftarrow c'$
        \STATE Resort remaining uncolored nodes by updated constraint level and uncertainty
      \ENDIF
    \ENDFOR
    \RETURN $C$
  \end{algorithmic}
\end{algorithm}

\begin{figure}[t!]
  \centering
  \includegraphics[width=0.8\linewidth]{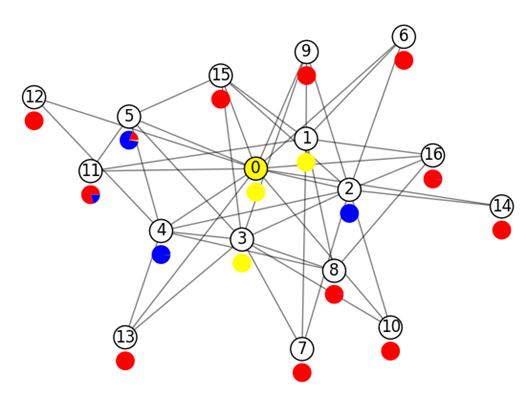}
  \caption{Graph representation showing predicted color probability distributions as pie charts for each node. Single-color pie charts denote confident model predictions, whereas multi-colored charts represent uncertain predictions. The GNN-heuristic approach utilizes these probability distributions to guide node color assignment decisions. Prediction uncertainty is demonstrated in nodes 5 and 11.}
  \label{fig:predicted_nodes}
\end{figure}

Model uncertainty can arise when nodes have identical local structures yet require distinct color assignments. For example, two central nodes connected to identical subgraphs might appear indistinguishable in their local context, but must still be assigned different colors. Similar ambiguities extend to leaf nodes and more complex graph symmetries. Conventional solutions would require more training data to reduce uncertainty. Instead, we leverage this ambiguity as a guiding signal for refinement opportunities. 

In particular, we algorithmically iterate over nodes and their predicted color distributions. Figure \ref{fig:predicted_nodes} illustrates an example of predicted color distributions for each node, visually represented as pie charts. 
  For nodes exhibiting uncertainty (indicated by multiple high-probability valid colors), we select the highest-probability color that satisfies all constraints and dynamically reorder nodes based on their available valid colors and connectivity (node degree). 

\subsection{Correction Stage with Constraint Programming Solver}\label{app:constraint}

As a fallback mechanism to guarantee solution validity, we use  a Constraint Satisfaction Problem (CSP) solver when preceding methods fail to produce a valid coloring. This stage is essential because the learning-based agent may occasionally generate colorings with constraint violations. 

The single-iteration refinement approach outlined in Algorithm~\ref{alg:passN_refinement} does not guarantee convergence to a valid solution, as demonstrated in Line 10 where all available colors may become unsafe for assignment. While our empirical results demonstrate that a properly fine-tuned GNN mitigates this limitation in practice, the potential for failure remains. Although incorporating backtracking mechanisms would provide theoretical guarantees for solution discovery when feasible, we adopt the CSP approach for its deterministic validation properties.

The CSP solver serves a dual purpose in our framework. First, it provides definitive validation by distinguishing between cases where conflicts arise from fundamental graph uncolorability versus limitations in the agent's decision-making pipeline. Second, this validation step enhances system explainability and trustworthiness by providing clear diagnostic feedback that separates algorithmic limitations from inherent problem constraints.

The CSP operates under two types of constraints:
\begin{itemize}
    \item \textit{\textbf{Hard constraints}}: Enforcing strict coloring rules to satisfy edge constraints
    \begin{equation}
        x_u \neq x_v,\quad \forall (u,v)\in E
    \end{equation}

    \item \textit{\textbf{Soft constraints}}: Minimizing deviations from the initial assignment generated by the GNN
    \begin{equation}
        \min_{x}\;\sum_{v\in V}\bigl|x_v - x_v^0\bigr|
    \end{equation}

\end{itemize}  

An additional balancing constraint can also be introduced, encouraging uniform color distribution:
\begin{equation}
        \min_{x}\;\sum_{c=1}^{C}\Biggl|\;\sum_{v \in V}\mathbf{1}[x_v=c] - \frac{|V|}{C}\;\Biggr|,
\end{equation}

which minimizes the deviation of each color class from the ideal balanced size \(|V|/C\). While this improves fairness across colors, in practice we found that it substantially increases runtime for larger graphs. Given our emphasis on efficiency, we exclude this term from the main pipeline but note its potential use in small-scale or diagnostic settings.

\subsection{Final Balancing Stage}
Finally, we conduct simulated annealing step after the coloring refinement steps. Nodes can adjust their colors provided that no violations are introduced. 
 Changes that improve balance are accepted, while less favorable adjustments are probabilistically retained or rejected based on a temperature-controlled acceptance criterion. This step continues iteratively until no further improvements are attainable or we reach a set limit (that we set for the number of iterations (We set the number of iterations to be equal to the number of nodes).

%% file: exp.tex
In this section, we first present validation experiments to justify our design decisions. In \textbf{Section A}, we investigate the training strategy by incorporating both the coloring loss ($L_{1}$) and the balancing loss ($L_{2}$). \textbf{Section B} examines the effect of varying the number of iterative refinements, thereby characterizing the standalone performance limits of the graph neural network which achieves solve rates between 70.83\% and 86.10\% on the test set and training set respectively. Next, in \textbf{Section C} we perform ablation studies to evaluate the different components of the workflow and their effect. Lastly, in \textbf{Section D}, we evaluate the same model on never‐seen‐before data and demonstrate a significant increase in solve rate due to the integration of Algorithm \ref{alg:passN_refinement}

\subsection{Balancing Coloring and Balancing Losses}  
As outlined in the methodology, the model is trained to focus on coloring loss (\( L1 \)) while optimizing balancing loss (\( L2 \)) as a secondary objective. We experiment with various techniques for merging these losses: 

\begin{itemize}
    \item \textbf{Baseline:} Model trained without \( L2 \).
    \item \textbf{Dynamic Balancing:} Adjusts \( \beta \) in \( L = L1 + \beta L2 \) based on \( L1 \). 
    \item \textbf{One Optimizer with Gradient Editing:} Computes gradients separately for \( L1 \) and \( L2 \), then combines them as \( d(L1)/d(\text{model}) + \beta d(L2)/d(\text{model}) \).
    \item \textbf{Two Gradients:} Processes the same batch twice per training step, updating model weights with two optimizers using different learning rates.
\end{itemize}

A simple model was trained on a per-instance basis, utilizing \( L1 \) and \( L2 \) losses across a dataset of 608 graphs. The performance results are presented in Table 1. Dynamic Balancing emerged as the top performer. The primary objective of this stage is to obtain a high-quality coloring solution, which it achieved by generating optimal solutions for 91.61\% of the training data and maintaining the lowest balancing error. Furthermore, it is important to note that for all these graph instances, passing the solution to a subsequent refinement stage resulted in a 100\% solve rate.

\begin{table}[ht]
\centering
\caption{Comparison of Different Loss Merging Techniques}
\label{tab:model_comparison}
\scalebox{0.8}{
\begin{tabular}{|l|c|c|c|}
\hline
 \textbf{Technique used}                      & \textbf{Solve rate} & \textbf{Mean Balance Loss} & \textbf{STD Balance Loss} \\ \hline
 Baseline                                   & 94.5\%                                         & 3.70                       & 3.32                      \\ \hline
\textbf{Dynamic Balancing }                                            & 91.61\%                                        & 1.06                       & 0.92                      \\ \hline                
One Optimizer with Gradient Editing                          & 91.9\%                                         & 1.32                       & 0.96                      \\ \hline
Two Optimizers                                                & 94.5\%                                         & 3.29                       & 2.85                      \\ \hline
\end{tabular}}
\end{table}

\subsection{Advantages of Iterative Refinement}
The present experiment aims to demonstrate the influence of the number of iterative refinement steps, specifically the number of forward passes, on the model's performance. As outlined in the methodology section, the colors produced by the last forward pass are re-introduced into the model as the initial coloring vector for the next iteration. This mechanism ensures that each subsequent pass benefits from a progressively improved starting point for the nodes, informed by the series of message passing operations already performed.
Table~\ref{tab:N_compare} compares performance across different values of number of forward passes. A single forward pass leads to poor results due to the two-stage training procedure, where the second stage is specifically designed to refine solutions that are already close to optimal. Although previous experiments showed that a single pass achieves a similar overall accuracy, the solutions exhibit poor balance, further highlighting the necessity of iterative refinement.

\begin{table}[ht]
\centering
\caption{Effect of the Number of Forward Passes of the Solving Rate on the In-house Dataset (GNN Model Only; Refinement Achieves 100\%)}
\label{tab:N_compare}
\begin{tabular}{|l|c|c|c|c|}
\hline
 \textbf{Number of Forward Passes}                      & \textbf{1} & \textbf{3} & \textbf{7} & \textbf{10} \\ \hline
Training set                                            & 23.7\%                                        & 76.6\%                       & 86.1\%     & 86.1\%                 \\ \hline              
Testing set                                             & 41.67\%                                         & 67\%                       & 70.83\%             & 70.83\%        \\ \hline
\end{tabular}
\end{table}

\begin{table*}[!t]
\centering
\caption{Comparison of Graph Coloring Algorithms}
\label{tab:algorithm_performance}
\begin{tabular}{|l|c|c|c|}
\hline
\textbf{Algorithm} & \textbf{Solve \%} & \textbf{Avg. Error} & \textbf{Std. Error} \\
\hline
\multicolumn{4}{|c|}{\textbf{Classical Baselines}} \\
\hline
DSATUR & 100\% & 2.31 & 1.30 \\
Welsh Powell & 91.71\% & 2.16 & 1.43 \\
\hline
\multicolumn{4}{|c|}{\textbf{Ours: Using Potts Loss only}} \\
\hline
Ours: Potts & 85.54\% & 2.07 & 1.10 \\
Ours: Potts + Algorithm \ref{alg:passN_refinement} & 100\% & 1.97 & 1.13 \\
Ours: Potts + Algorithm \ref{alg:passN_refinement} + SA & 100\% & 1.94 & 1.05 \\
\hline
\multicolumn{4}{|c|}{\textbf{Ours: Adding the Balancing Loss}} \\
\hline
Ours: Potts + Balance & 83.64\% & 1.86 & 0.976 \\
Ours: Potts + Balance + Algorithm \ref{alg:passN_refinement} & 99.76\% & 1.83 & 0.972 \\
Ours: Potts + Balance + Algorithm \ref{alg:passN_refinement} + SA & 99.76\% & 1.81 & 0.928 \\
\hline
\multicolumn{4}{|c|}{\textbf{Ours: Adding Entropy Loss}} \\
\hline
Ours: Potts + Balance + Entropy & 82.46\% & 2.03 & 1.06 \\
Ours: Potts + Balance + Entropy + Algorithm \ref{alg:passN_refinement} & 100\% & 1.92 & 1.02 \\
Ours: Potts + Balance + Entropy + Algorithm \ref{alg:passN_refinement} + SA & 100\% & 1.90 & 0.99 \\
\hline
\multicolumn{4}{|c|}{\textbf{Ours: Substituting the Balancing Loss with a Stricter Variant}} \\
\hline
Ours: Potts + Harsh Balance + Entropy & 77.06\% & 1.74 & 0.89 \\
Ours: Potts + Harsh Balance + Entropy + Algorithm \ref{alg:passN_refinement} & 100\% & 1.75 & 0.90 \\
Ours: Potts + Harsh Balance + Entropy + Algorithm \ref{alg:passN_refinement} + SA & 100\% & 1.716 & 0.90 \\
\hline
\end{tabular}
\end{table*}

\subsection{Ablation Study on Graph Coloring with Balance and Entropy Regularization}

To evaluate the contributions of different model components in our GNN-based framework for graph coloring, we conduct an extensive ablation study summarized in Table~\ref{tab:algorithm_performance}. We compare classical heuristics (DSATUR and Welsh Powell) against several GNN variants trained with combinations of Potts loss, balance regularization, entropy minimization with GNN-heuristic and simulated annealing.

In this experiment, the model is trained exclusively using the second stage (Dynamic Balancing) for 500 epochs. The results are reported on the 432 training graphs.

\paragraph{Baselines.} Classical heuristics achieve high solve rates (DSATUR: 100\%, Welsh Powell: 91.71\%) but exhibit significantly higher average balancing errors (2.31 and 2.16, respectively) and larger variability (std.\ error of 1.3 and 1.43). These results suggest that while traditional methods are feasible in terms of coloring, they are not well-suited for producing balanced solutions.

\paragraph{Effect of Potts Loss Alone.} Using a GNN trained solely with Potts loss (\texttt{Ours: Potts}) reduces average error to 2.07, outperforming the classical baselines. Introducing Algorithm \ref{alg:passN_refinement} and simulated annealing (\texttt{\_order\_SA}) improves solve percentage to 100\% and further lowers avg.\ error to 1.97 and 1.94, respectively. This highlights the benefit of an algorithm-based stage for enhancing both feasibility (Algorithm \ref{alg:passN_refinement}) and balance (Simulated Annealing).

\paragraph{Impact of Balance Regularization.} Incorporating balance loss (\texttt{\_Balance}) has a significant effect on solution quality. Without any Algorithm \ref{alg:passN_refinement}, the avg.\ error drops to 1.86, albeit with a lower solve rate (83.64\%). Adding Algorithm \ref{alg:passN_refinement} and simulated annealing improves both metrics, achieving 100\% solve rate and reducing avg.\ error to 1.83 and 1.81, respectively. This confirms the effectiveness of explicitly regularizing the color distribution as part of the GNN's loss function.

\paragraph{Adding Entropy Minimization.} Entropy regularization alone (\texttt{\_Balance\&Entropy}) does not consistently outperform balance-only variants (avg.\ error: 2.03; solve rate: 82.46\%). However, when combined with Algorithm \ref{alg:passN_refinement} (\texttt{\_Balance\&Entropy\_order}) and annealing (\texttt{\_Balance\&Entropy\_order\_SA}), performance improves significantly, achieving 100\% solve and lowering the avg.\ error to 1.92 and 1.90 with reduced variance. This suggests that entropy aids in model stability and generalization when used jointly with other losses.

\paragraph{Harsh Balance for Stronger Regularization.} A more aggressive balance constraint (\texttt{HarshBalance}) results in lower avg.\ error (1.74) but reduced solve rate (77.06\%) when entropy is also used. Notably, when augmented with Algorithm \ref{alg:passN_refinement} and SA, the model achieves 100\% solve rate and the lowest observed avg.\ error (1.716) with low variance (std.\ error: 0.9). This indicates that harsher balancing improves solution quality, but only when adequately supported by scheduling strategies.

\paragraph{Key Takeaways.} 
\begin{itemize}
    \item Balance regularization substantially improves error metrics, outperforming classical baselines and pure Potts-based learning.
    \item Entropy minimization aids solution stability when used in conjunction with other objectives.
    \item Scheduling techniques such as Algorithm \ref{alg:passN_refinement} and annealing are crucial for ensuring high feasibility (100\% solve rate).
    \item The optimal configuration combines Potts, Harsh balance and entropy losses and the algorithmic refinement stage, achieving the lowest balancing error and full feasibility.
\end{itemize}

This ablation confirms that careful integration of balance, entropy, and scheduling leads to state-of-the-art performance in both feasibility and solution quality.

\subsection{Benchmarking results on OpenMPL data}
As previously mentioned, we leverage the OpenMPL framework to generate reproducible graph instances from benchmark datasets. Table \ref{tab:optimized_comparison} presents a comparative evaluation of our proposed pipeline (GNN with Pass@N correction) against the OpenMPL binaries invoked with the \verb|-algo ILP\_UPDATED| option. Our pipeline achieves identical solving rates but improves balancing.

\begin{table}[ht]
\centering
\caption{Comparison of our proposed pipeline vs. OpenMPL's \textit{-algo ILP\_UPDATED} approach on two benchmark instances.}
\label{tab:optimized_comparison}
\scalebox{0.85}{
\begin{tabular}{|l|c|c|}
\hline
 \multicolumn{3}{|c|}{\textbf{sim\_c1.gds}} \\ \hline
Metric               & \textbf{GNN with Algorithm \ref{alg:passN_refinement} correction} & \textbf{ILP\_updated} \\ \hline
Conflict-count       &          4            &                4              \\ 
Color1 count         &          370             &       369                        \\ 
Color2 count         &          369             &       373                        \\ 
Color3 count         &          370            &        367                       \\ \hline
 \multicolumn{3}{|c|}{\textbf{sim\_c2.gds}} \\ \hline
Metric               & \textbf{GNN with Algorithm \ref{alg:passN_refinement} correction} & \textbf{ILP\_updated} \\ \hline
Conflict-count       &          0             &               0                \\ 
Color1 count         &          739            &                 746             \\ 
Color2 count         &          740            &                 728              \\ 
Color3 count         &          740          &                 742              \\ \hline
\end{tabular}}
\vspace{-10pt}
\end{table}

\subsection{Discussion of Experimental Results}

This section summarizes key insights, observations, and practical considerations derived from the experimental results.

First, while our approach may initially seem more involved than traditional solvers, it offers a critical benefit: the GNN is capable of producing predictions in constant time, unlike purely algorithmic methods, which typically require iterative procedures. As shown in Table~\ref{tab:algorithm_performance}, the quality of the initial solution has a significant impact on the overall performance. For example, comparing the variants \textit{Ours: Potts only} versus \textit{Potts + Harsh Balancing + Entropy}, we observe that the algorithmic refinement stages benefit substantially from stronger initialization.

Additionally, unlike some prior work that trains a separate model per instance, we utilize the fact that the graphs in our setting are not arbitrary; they represent real circuit structures and therefore contain recurring, meaningful patterns. The GNN is designed to absorb and exploit these patterns effectively. This stands in contrast to other optimization problems, where the input graphs may be random or lack any inherent structure. In such cases, there is often nothing consistent for the model to learn from. Here, we know that the graphs carry domain-specific properties, which gives the model a useful signal to generalize across instances without retraining from scratch.

Training the model for 500 epochs requires only a few minutes on an NVIDIA v100 GPU, highlighting the computational efficiency of the approach. This efficiency suggests that the proposed method can be adopted in practice at low cost and can serve as a strong initialization method, improving upon the commonly used random or heuristic-based starting solutions in industrial applications.

While the graphs used in our experiments may appear small, it is important to note that they represent the most complex components of the problem space. The graphs have already been reduced using existing algorithmic pre-processing techniques, as described in the introduction. The model is thus tasked with handling the most interconnected, constraint-heavy subgraphs, where algorithmic methods tend to struggle the most.

%% file: future.tex
We identify three complementary research avenues that address fundamental limitations in current refinement-based optimization frameworks.

\subsection{Learnable Refinement Architectures}
Current heuristic-based refinement methods lack adaptability to problem-specific structures. We propose developing end-to-end trainable refinement modules that can learn domain-specific improvement strategies.We plan to investigate reinforcement learning that can be jointly optimized with the initial solution generator. This approach promises to move beyond hand-crafted heuristics toward learned, generalizable refinement policies.

\subsection{Reasoning-Enhanced Optimization via Large Language Models}
The emergence of sophisticated reasoning capabilities in modern LLMs presents an opportunity to enhance both refinement quality and overall pipeline intelligence. We intend to explore two complementary approaches: (1) leveraging reasoning models as intelligent refinement operators that can perform multi-step logical improvements on candidate solutions, and (2) developing agentic frameworks where LLMs orchestrate the entire optimization pipeline, dynamically selecting refinement strategies based on problem characteristics and intermediate results. 

\subsection{Advanced Balancing Mechanisms}
While simulated annealing provides a principled approach to exploration-exploitation trade-offs, its performance is often limited by parameter sensitivity and problem-agnostic design. We propose investigating more sophisticated final stage balancing mechanisms.
Each direction will be evaluated using comprehensive benchmarks spanning diverse problem domains, with particular attention to scalability, generalization, and computational efficiency trade-offs.

%% file: conc.tex
In this paper, we proposed a GNN methodology for constrained graph coloring tailored specifically to multipatterning problems in EDA. Our approach leverages an unsupervised GNN agent to autonomously uncover inherent graph constraints and structural properties and generate initial coloring predictions that progressively reduce violations through a built-in, self-correcting mechanism. Any residual inconsistencies are subsequently refined and corrected  through multiple iterative stages. Our experiments demonstrated robust solutions that effectively reduced constraint violations while balancing color distributions. This multi-stage framework proved adaptable to the complex constraints typical of EDA layout.